\documentclass[preprintnumbers,nofootinbib,onecolumn,floatfix,letterpaper,superscriptaddress,11pt]{revtex4}
\pdfoutput=1
\usepackage{epsfig}
\usepackage{amsfonts}
\usepackage{graphicx}
\usepackage{epsfig}
\usepackage{eepic}
\usepackage{amsmath}
\usepackage{amssymb}
\usepackage{color}
\usepackage{dcolumn}% Align table columns on decimal point
\usepackage{bm}% bold math
\usepackage{ulem}
\usepackage{slashbox}

\def\bc{\begin{center}}
\def\nno{\nonumber}
\def\ec{\end{center}}
\def\be{\begin{eqnarray}}
\def\ee{\end{eqnarray}}
\definecolor{dyellow}{rgb}{1.,0.8,.0}
\definecolor{myblue}{rgb}{.1,.1,.7}
\definecolor{dcyan}{rgb}{.0,.6,.6}
\definecolor{dmagenta}{rgb}{0.6,0.0,0.6}
\definecolor{brown}{rgb}{0.6,0.2,0.}
\definecolor{darkblue}{rgb}{.0,.0,0.5}
\definecolor{darkred}{rgb}{0.75,0.0,0.0}
\definecolor{orange}{rgb}{1.,.6,.0}
\definecolor{dorange}{rgb}{0.8,.4,.0}
\definecolor{darkgreen}{rgb}{0.0,0.6,0.0}
\definecolor{purple}{rgb}{.4,.0,.4}
\definecolor{lightgrey}{rgb}{0.7, 0.7, 0.7}
\definecolor{grey}{rgb}{0.4, 0.4, 0.4}
\def\pa{\partial}

\newcommand{\nc}{\newcommand}
\nc{\rnc}{\renewcommand} \nc{\ket}[1]{\left | \, #1 \right \rangle}
\nc{\bra}[1]{\left \langle #1 \, \right |}
\nc{\ua}{\uparrow} \nc{\da}{\downarrow}

\nc{\braket}[2]{\langle\, #1\,|\,#2\,\rangle}
\nc{\half}{\frac{1}{2}}

\nc{\prj}{\mathcal{P}} \nc{\hilb}{\mathcal{H}}
\nc{\pth}{\mathcal{C}} \nc{\inprod}[2]{\braket{#1}{#2}}
\nc{\upket}{\ket{\uparrow}} \nc{\downket}{\ket{\downarrow}}
\nc{\upbra}{\bra{\uparrow}} \nc{\downbra}{\bra{\downarrow}}

\begin{document}
%\begin{CJK*}{GBK}{song}

\title{A holographic model of SQUID}

\author{Rong-Gen Cai}\email{cairg@itp.ac.cn}
\affiliation{State Key Laboratory of Theoretical Physics, Institute of Theoretical Physics,
Chinese Academy of Sciences, Beijing 100190, China;}

\author{Yong-Qiang Wang}\email{yqwang@lzu.edu.cn}
\affiliation{Institute of Theoretical Physics, Lanzhou University, Lanzhou 730000, China;}

\author{Hai-Qing Zhang}\email{hqzhang@cfif.ist.utl.pt}
\affiliation{CFIF, Departamento de F\'isica, Instituto Superior {T}\'ecnico,
Universidade {T}\'ecnica de Lisboa, Av. Rovisco Pais 1, 1049-001
Lisboa, Portugal.}

\begin{abstract}
We construct a holographic model of superconducting quantum interference device (SQUID) in the Einstein-Maxwell-complex scalar theory with a negative cosmological constant. The SQUID ring consists of two Josephson junctions which sit on two sides of a compactified spatial direction of a Schwarzschild-AdS black brane.  These two junctions interfere with each other and then result in a total current depending on the magnetic flux, which can be deduced from the phase differences of the two Josephson junctions. The relation between the total current and the magnetic flux is obtained numerically.
\end{abstract}

 \maketitle
\section{Introduction}
The  anti-de Sitter (AdS)/conformal field theory (CFT) correspondence~\cite{Maldacena:1997re,Gubser:1998bc,Witten:1998qj} relates a weakly coupled gravity in AdS space to a strongly coupled CFT in a lower dimension. On of the active arenas of its applications is condensed matter physics.
In recent years, some important progresses have been made in this area.  For example, some gravitational dual models of superfluid/superconductor~\cite{Gubser,Hartnoll:2008vx}, (non-)Fermi liquid~\cite{Lee:2008xf,Liu:2009dm,Cubrovic:2009ye}, and Josephson junctions~\cite{Horowitz:2011dz,Kiritsis:2011zq} have been constructed and intensively studied. For recent reviews, please refer to \cite{Hartnoll:2009sz,Lecture2,Iqbal:2011ae}.

\begin{figure}[h]
 \includegraphics[trim=0cm 1.cm 0cm 17cm, clip=true, scale=0.4,angle=0]{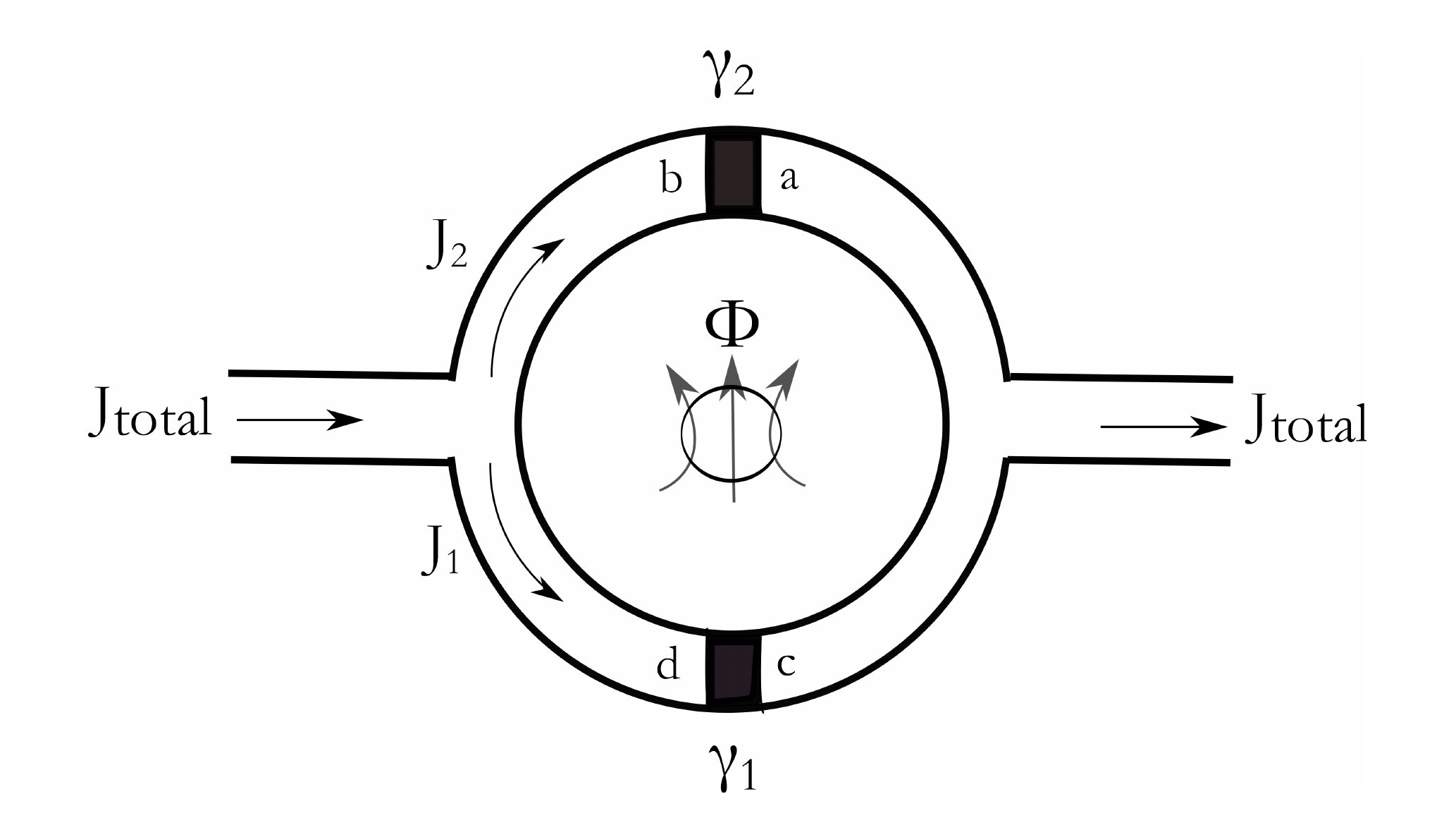}
   \includegraphics[trim=0cm 0cm 0cm 0cm, clip=true, scale=0.5]{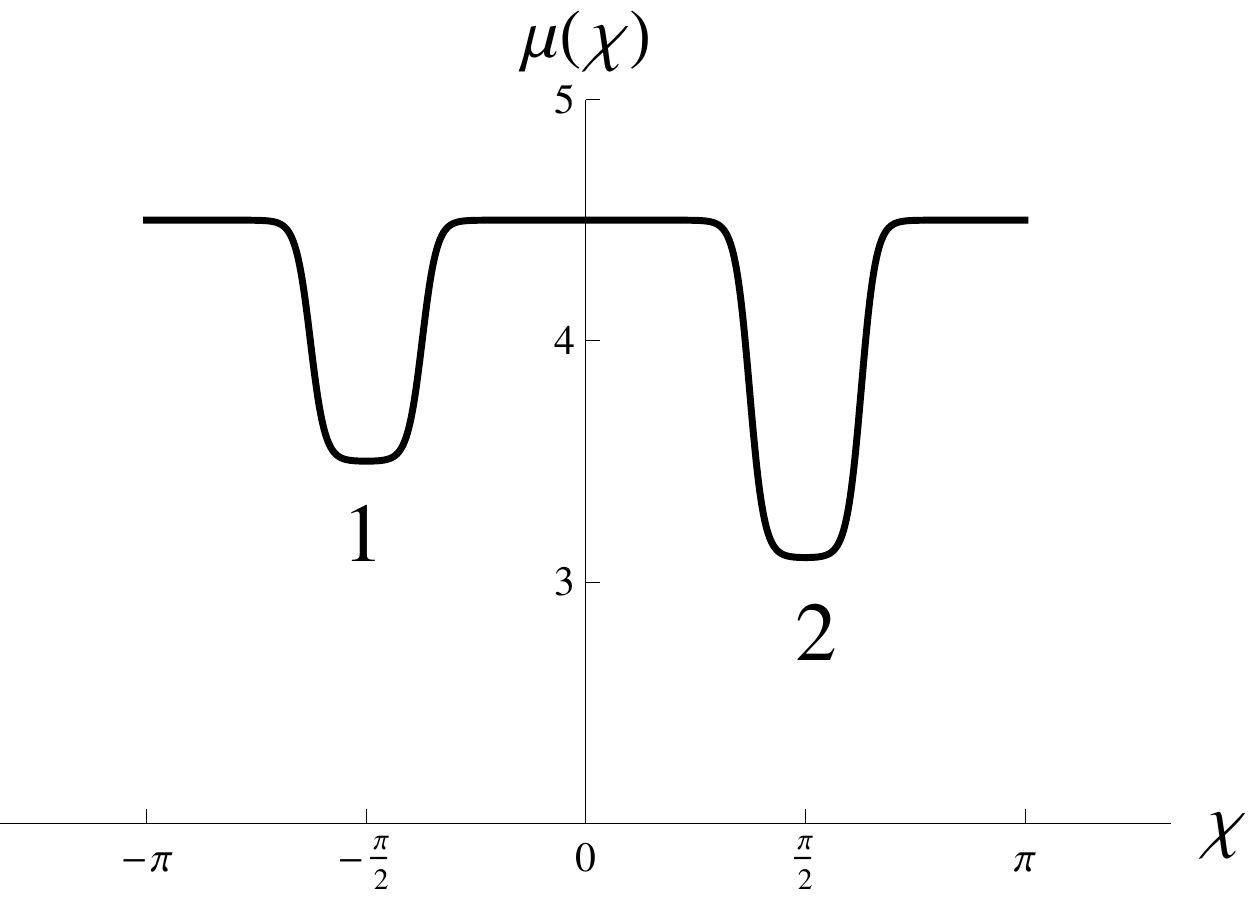}
  \caption{\label{mu} (Left.) A schematic cartoon of the SQUID in condensed matter physics. $\Phi$ is the magnetic flux  through the ring. The black parts $d\to c$ and $b\to a$ are the two junctions 1 and 2, respectively; (Right.) The chemical potential $\mu(\chi)$ along the SQUID ring. The number $1$ and $2$ represent the junction $1$ and junction $2$  in the left panel, respectively. }
\end{figure}

As an important practical application of superconductivity, superconducting quantum interference device (SQUID)~\cite{tinkham} can  detect extremely weak magnetic field strength.  The SQUID is a superconducting ring in which there are two Josephson junctions~\cite{Josephson:1962zz} sitting on two sides of the ring. A schematic cartoon of the SQUID ring is plotted on the left panel of Fig.\ref{mu}. These two Josephson junctions will interfere with each other, and then the total current $J_{total}$ will depend on the phase differences of the two junctions. Furthermore, the net difference of the two phase differences is proportional to the magnetic flux $\Phi$  through the SQUID ring. The relation between the total current and the magnetic flux is~\cite{tinkham}
\be\label{j1j2}
J_{total}&=&J_{1c}\sin(\gamma_1)+J_{2c}\sin(\gamma_2)\nno\\
%&=&2J_{1c}\cos\left(\frac{\gamma_2-\gamma_1}{2}\right)\sin\left(\frac{\gamma_1+\gamma_2}{2}\right)\nno\\
&=&2J_{1c}\cos\left(\frac{\Phi}{2}\right)\sin(\gamma),
\ee
 where $J_{1c}=J_{2c}$ is assumed and they are the maximal currents of two Josephson junctions;  $\gamma_1$ and $\gamma_2$ are respectively the phase differences of junction 1 and 2,  while  $\gamma=(\gamma_1+\gamma_2)/2+\pi n$ and $\Phi=(\gamma_2-\gamma_1)+2\pi n$ ($n$ is an integer, and can be referred to as the fluxoid number~\cite{tinkham}). The magnetic flux $\Phi$ can be obtained through the integration of the gauge field along the ring. See also \cite{Montull:2011im,Montull:2012fy,Cai} for the relation between $\Phi$ and the integration of gauge field along a compactified direction in a holographic setup.

In this paper,  we will construct a holographic model of the SQUID ring by putting it in a compactified spatial direction $\chi$ on the boundary of  a Schwarzschild-AdS black brane. We will choose a specific type of chemical potential which can model the superconductor-normal metal-superconductor (SNS) Josephson junction on two sides of the ring, please refer to the right panel of Fig.\ref{mu}.
 Instead of obtaining the usual relation \eqref{j1j2} in condensed matter physics, we actually get a more general form for the sake of the numerical calculation convenience.

 The paper is arranged as follows: In Sec.\ref{sect:model}, we construct the model in a Schwarzschild-AdS black brane with a specific chemical potential; Numerical results are shown in Sec.\ref{sect:num}; Finally, we draw the conclusions and discussions in Sec.\ref{sect:con}.

 \section{Setup of the model}
 \label{sect:model}
 The matter sector of the model is described by the Maxwell-complex scalar theory as
 \be S=\int d^4x\sqrt{-g}(-\frac14 F_{\mu\nu}F^{\mu\nu}-|\nabla\psi-iA\psi|^2-m^2|\psi|^2),\ee
  where $A_\mu$ is the U(1) gauge field while $F_{\mu\nu}=\pa_\mu A_\nu-\pa_\nu A_\mu$ is the field strength. We will work in the probe limit, namely the back reaction of matter fields on the background geometry will be neglected.  The gravitational background is a $(3+1)$-dimensional Schwarzschild-AdS black brane given by (we have scaled the AdS radius $L\equiv1$)
 \be \label{metric} ds^2=-f(r)dt^2+\frac{1}{f(r)}dr^2+r^2(dx^2+d\chi^2).\ee
 where $f(r)=r^2-r_0^3/r$ with $r_0$ the horizon radius of the black brane. The temperature of the black brane is $T=3r_0/(4\pi)$. The direction $\chi$ is compactified with the periodicity $-\pi R\leq \chi\leq\pi R$ in which $R$ is the radius of the $\chi$-loop. The gravitational background \eqref{metric} is thermodynamically favored  when $T>1/(2\pi R)$ \cite{Montull:2012fy}. Therefore in the following we will set $r_0=R=1$, which satisfies the above condition.
 It is convenient for us to choose a gauge for the matter fields as
 \be \psi=|\psi|e^{i\phi},\quad A_\mu=(A_t,~A_r,~0,~A_\chi),\ee
 where $|\psi|, \phi, A_t, A_r,$ and $A_\chi$ are all real functions of $r$ and $\chi$. In the following context we will work with the gauge-invariant quantity $M_\mu \equiv A_\mu-\partial_\mu\phi$.
 The equations of motion (EoMs) of the matter sector in the background (\ref{metric})  are
% \lipsum[1]
\begin{widetext}
 \begin{subequations}
 \begin{align}
 \frac{\pa_\chi^2{M_t}}{r^2 f}-\frac{2 {M_t} |\psi|
   ^2}{f}+\frac{2 \pa_r{M_t}}{r}+\pa_r^2{M_t}&=0,\label{eom1}\\
 -\pa_\chi^2{M_r}+2 r^2 {M_r} |\psi| ^2+\pa_{r\chi}{M_\chi}&=0,\label{eom2}\\
 -\frac{f' \pa_\chi{M_r}}{f}+\frac{f' \pa_r{M_\chi}-2
   {M_\chi} |\psi| ^2}{f}-\pa_{r\chi}{M_r}+\pa_r^2{M_\chi}&=0,\label{eom3}\\
 \left(\frac{f'}{f}+\frac{2}{r}\right) \pa_r|\psi| -\frac{{m^2} |\psi|
   }{f}+\frac{{M_t}^2 |\psi| }{f^2}-\frac{{M_\chi}^2
   |\psi| }{r^2 f}+\frac{\pa_\chi^2|\psi| }{r^2 f}-{M_r}^2 |\psi|
   +\pa_r^2|\psi| &=0,\label{eom4}\\
 f' {M_r} |\psi| +f |\psi|\pa_r{M_r}  +2 f
   {M_r} \pa_r|\psi| +\frac{2 f {M_r} |\psi|
   }{r}+\frac{|\psi|\pa_\chi{M_\chi}  }{r^2}+\frac{2 {M_\chi}
   \pa_\chi|\psi| }{r^2}&=0.\label{eom5}
 \end{align}
 \end{subequations}
 \end{widetext}
  where $f'\equiv\pa_rf$. The above equations are not independent, in particular, one has ${\pa_r}(\text{Eq.}\eqref{eom2}\times f)-{\pa_\chi}(\text{Eq.}\eqref{eom3}\times f)-2r^2|\psi|\times \text{Eq.}\eqref{eom5} =0$.
 Therefore, there are four independent EoMs for four fields, {\it i.e.}, $|\psi|, M_t, M_r$ and $M_\chi$.

At the horizon $r=r_0$, the fields $M_t$ should be vanishing $M_t=0$ in order to make $g^{tt}M^2_t$ regular there, because $g^{tt}$ is divergent at the horizon. Other fields should be finite at the horizon.

At the AdS boundary $r\to\infty$, the asymptotic behaviors of the fields are of the forms
\be
|\psi|(r,\chi)&\sim&\frac{|\psi|^{(1)}(\chi)}{r^{(3-\sqrt{9+4m^2})/2}}+\frac{|\psi|^{(2)}(\chi)}{r^{(3+\sqrt{9+4m^2})/2}}+\cdots,\\
M_t(r,\chi)&\sim&\mu(\chi)-\frac{\rho(\chi)}{r}+\cdots,\\
M_r(r,\chi)&\sim& \frac{M_r^{(2)}(\chi)}{r^2}+\cdots,\\
\label{myexpansion}M_\chi(r,\chi)&\sim&\nu(\chi)+\frac{J(\chi)}{r}+\cdots.
\ee
From the AdS/CFT dictionary \cite{Gubser:1998bc,Witten:1998qj}, $|\psi|^{(1)}$ and $|\psi|^{(2)}$ can be regarded as the source and vacuum expectation value of the corresponding operator $O$ dual to the scalar field $|\psi|$. We here turn off the source term and therefore impose $|\psi|^{(1)}\equiv0$ in the following numerical calculations because we  require the U(1) symmetry to be spontaneously broken; $\mu$ and $\rho$ are the chemical potential and charge density of the dual field theory, respectively; While $\nu$ and $J$ can be interpreted as the superfluid velocity and current of the dual field theory.\footnote{In the expansions near the boundary, there is a term like ${\partial_\chi J}=2(|\psi|^{(1)})^2M_r^{(2)}-{\partial^2_\chi M_r^{(2)}}$.
Therefore, if we set $M_r^{(2)}(\chi)=0$ at the boundary, it is easy to infer that $J(\chi)$ should be a constant which is similar to the case in the literatures \cite{Horowitz:2011dz,Wang:2011rva,Wang:2011ri,Wang:2012yj}.}

Note that in the homogeneous case, {\it i.e.}, all the fields are independent of the coordinate $\chi$, the critical chemical potential at the superconductor/normal metal phase transition is $\mu_c\approx4.06$ for $m^2=-2$~\cite{Hartnoll:2008vx,Horowitz:2011dz}, and that
  a higher chemical potential corresponds to a lower temperature, and {\it vice versa}. Thus in the numerical calculations we can tune the chemical potential while fixing the temperature~\cite{Hartnoll:2008vx,Horowitz:2011dz}; This is equivalent to tune the temperature while fixing the chemical potential. In our numerical calculations we will set the chemical potential on the boundary as
\be
\label{chem}
 \mu(\chi)&=&h-\sum_{i=1,2}d_i\left[\tanh\left(\frac{k_i(\chi-p_i+w_i)}{\pi}\right)-\tanh\left(\frac{k_i(\chi-p_i-w_i)}{\pi}\right)\right],
\ee
in which $i=1,2$ stand for the junctions in the SQUID ring (see Fig.\ref{mu}), and $h, d_i, k_i, p_i$, and $w_i$ are related to the highest value, depth, slope, position, and width of the junction $i$, respectively.  Please see the right panel of Fig.\ref{mu} for a typical chemical potential in our model, we have set that only the depths of the two junctions are distinct. In this plot, the parameters we choose are $h=4.5$, $(d_1, d_2)=(0.5, 0.7)$, $(k_1, k_2)=(30, 30)$, $(p_1, p_2)=(-\pi/2, \pi/2)$ and $(w_1, w_2)=(0.4, 0.4)$, respectively.
We can see from Fig.\ref{mu} that the higher parts of the chemical potential are greater than $\mu_c$, therefore, these parts correspond to the superconductors, while the lower parts corresponding to the normal metals are smaller than $\mu_c$. Therefore, when applying this chemical potential on the compactified $\chi$-loop, we can realize the holographic model of a SQUID on the boundary.

%\subsection{An simple example of holographic realization}
\section{Numerical results}
\label{sect:num}
In the numerical calculations, we have scaled
\be
|\psi|\to\frac{|\psi|}{r^{{(3-\sqrt{9+4m^2})/2}}},\quad M_r\to\frac{M_r}{r^2}.
\ee
for numerical convenience. In addition, it is convenient to work in the $(z=1/r,\chi)$-coordinates, thus, $z=0$ now is the AdS boundary and $z=1$ is the horizon. We will work in the case with $m^2=-2$. The numerical methods we adopted are the combination with the Chebyshev spectral method and the Newton-Raphson method \cite{trefethn}.

\begin{figure}[h]
   \includegraphics[trim=1.0cm 8.5cm 1.5cm 8.5cm, clip=true, scale=0.43]{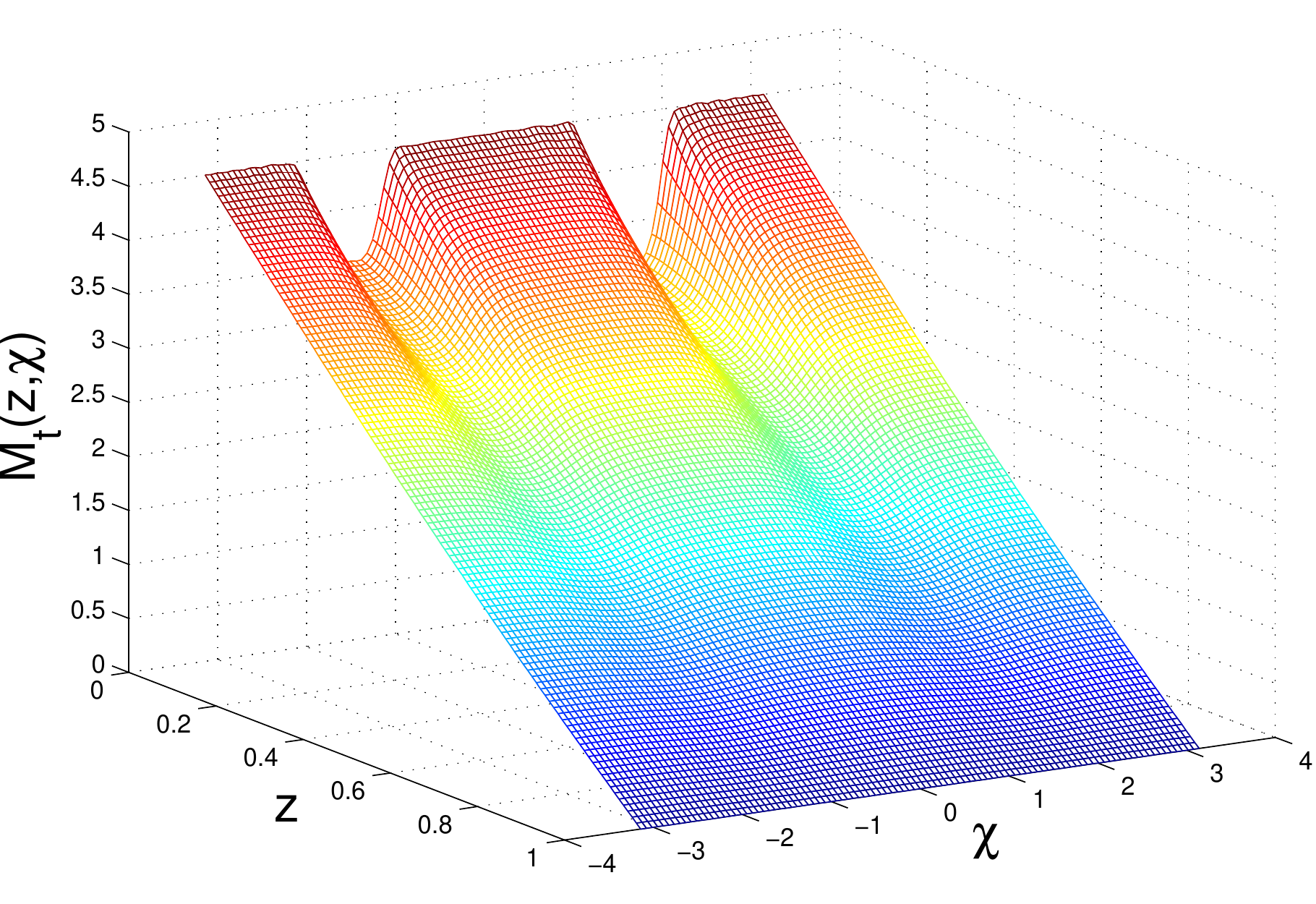}
   \includegraphics[trim=1.0cm 8.5cm 1.5cm 8.5cm, clip=true, scale=0.43]{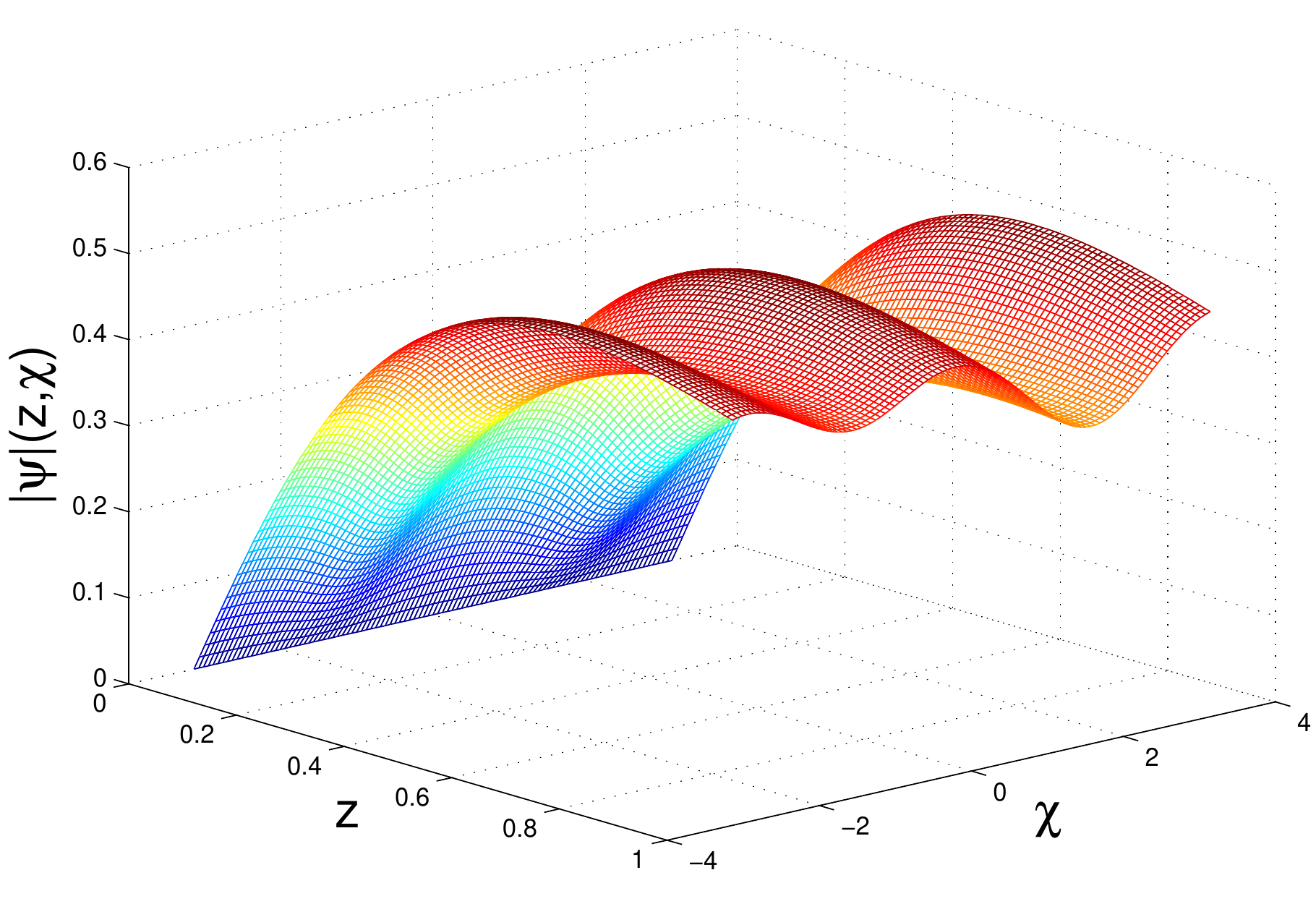}
   \includegraphics[trim=1.0cm 7.7cm 2.0cm 8.5cm, clip=true, scale=0.45]{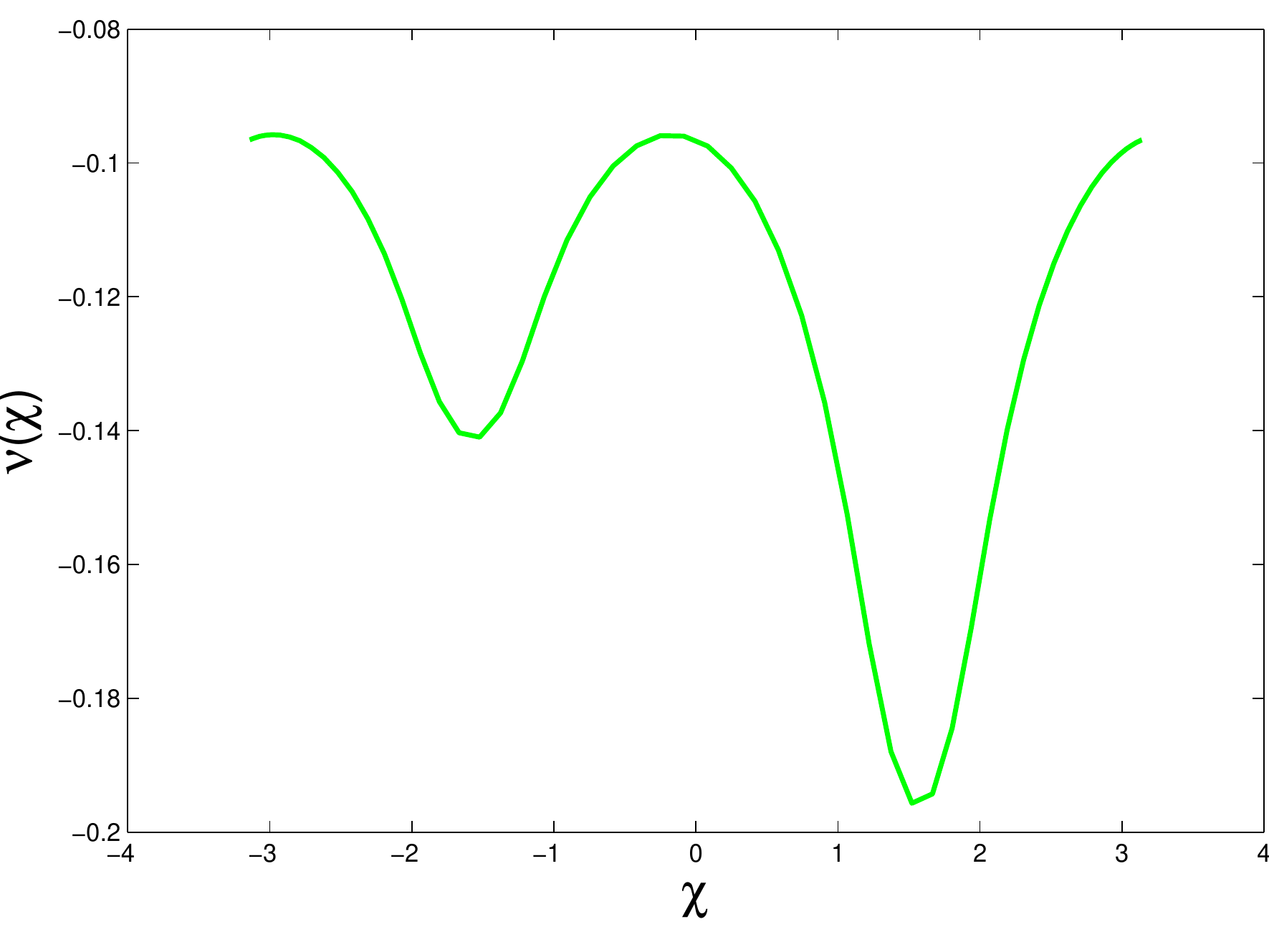}
    \includegraphics[trim=1cm 7.7cm 2cm 8.5cm, clip=true, scale=0.45]{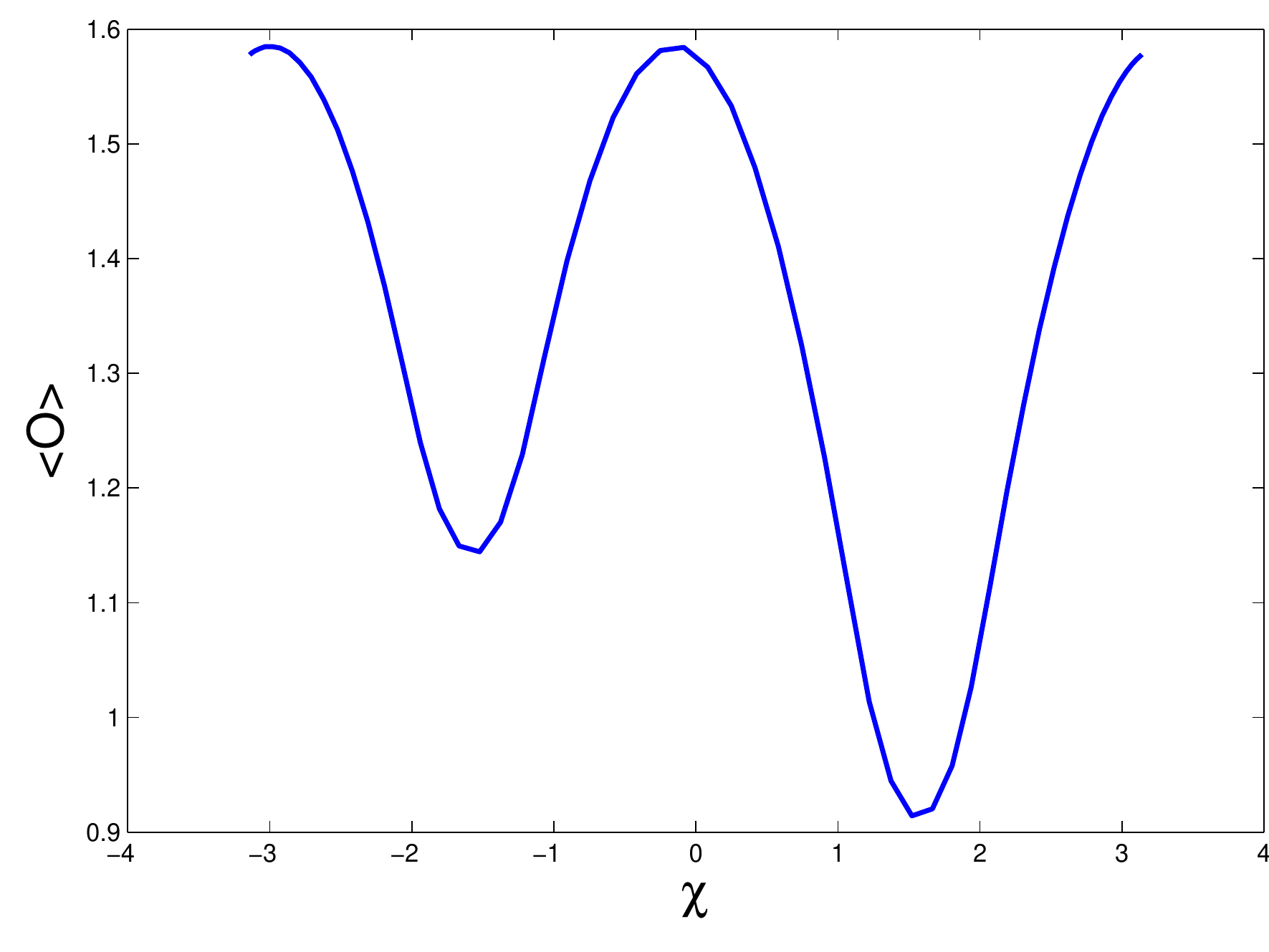}
  \caption{\label{fields}  The profiles of $M_t$, $|\psi|$, $\nu$ and $\langle O\rangle$. In these plots, the parameters are  $J=0.03$, $m^2=-2$, $h=4.5, (d_1, d_2)=(0.5, 0.7), (k_1, k_2)=(30, 30), (p_1, p_2)=(-\pi/2, \pi/2)$ and $(w_1, w_2)=(0.4, 0.4)$.}
\end{figure}

As shown in Fig.\ref{mu}, $J_1$ is the current flowing through the lower junction ($-\pi\to0$), while $J_2$ is the current flowing through the upper one ($\pi\to0$). Following \cite{Horowitz:2011dz}, we can solve the equations of motion \eqref{eom1}-\eqref{eom5} along the the lower junction and upper junction, viewing $J_1$ and $J_2$ as input parameters, respectively. However, to model a SQUID, we need to set $J_1=J_2$ in order to make the scalar field $|\psi|$ be continuous at the two ends of the lower and upper junctions, {\it i.e.,} $|\psi(\chi=\pi)|=|\psi(\chi=-\pi)|$ and $|\psi(\chi=0^+)|=|\psi(\chi=0^-)|$. The continuity of the scalar field at the two ends is crucial in deriving the formula \eqref{j1j2} in condensed matter physics~\cite{tinkham}. Therefore, in the numerical calculations we set the supercurrent $J_1=J_2=J=\text{constant}$ as the input parameter and impose the continuous conditions for $|\psi|$ at the two ends.  Of course, it is not necessary to impose the continuous conditions for other fields, such as $M_t$, $M_r$ and $M_\chi$, at the two ends. But, in practice these gauge fields are also continuous at the two ends because of the input parameters $J_1=J_2$ and the continuity of chemical potential we choose at the two ends. In the numerical calculations it is helpful to note that there is a symmetry in the equations of motion \eqref{eom1}-\eqref{eom5},
  \be M_\chi\to-M_\chi, M_r\to-M_r, M_t\to M_t, |\psi|\to |\psi|. \ee
 We now define the gauge-invariant phase difference for the junction 1 and 2 as
\be\label{gamma12} \gamma_1&=&-\int^0_{-\pi}\left(\nu(\chi)-\nu(0)\right)d\chi, \\
  \label{gamma122}\gamma_2&=&-\int_\pi^{0}\left(\nu(0)-\nu(\chi)\right)d\chi, \ee
 respectively, which are similar to the one in Ref.~\cite{Horowitz:2011dz}.  But in the definition of $\gamma_2$ we have added an extra minus. This is due to the fact that in the numerical calculations (see Fig.\ref{fields}) the integration from $\pi$ to $0$ will contribute an extra minus sign. As a result, the extra minus sign in the definition of $\gamma_2$ can cancel the effect of the minus sign coming from the
 numerical integration. In this way one can obtain the correct phase difference for the upper Josephson junction.  Thus we can model a SQUID in which the total current flows into from $\chi=\pm\pi$ and flows out from $\chi=0$ rather than a circuit current flowing around the loop.\footnote{ A simple check of the correctness of our definition for the phase difference is as follows:  If the chemical potentials for the two junctions are same,  the upper and lower junctions are then exactly identical. In this case, there does not exist any interference between these two junctions. This means $\Phi=\gamma_2-\gamma_1=0$, which can be exactly obtained from Eq.\eqref{gamma12} and Eq.\eqref{gamma122}.} So the total current flowing out from $\chi=0$ is $J_{total}=J_1+J_2=2J$. We plot the profiles of the fields $M_t, |\psi|, \nu$ and $\langle O\rangle$ in Fig.\ref{fields} with $J=0.03$.

Next we will  get the numerical results between the total current $J_{total}$ and the magnetic flux $\Phi$.
 In condensed matter physics~\cite{tinkham}, usually one demands that the maximal currents of the junctions on both sides are identical, namely, $J_{1c}=J_{2c}$. In that case, one can deduce the famous formula Eq.\eqref{j1j2} for the SQUID, and the maximum of the total current will depend on the magnetic flux as $J_c=2J_{1c}|\cos\left({\Phi}/{2}\right)|$.  In principle, one can obtain $J_{1c}=J_{2c}$ with different $\gamma_1$ and $\gamma_2$ by properly adjusting the parameters in the chemical potential (\ref{chem}). But in practice, it is quite difficult to arrive at this goal in order to satisfy the periodic condition for the scalar fields. On the other hand, in order to have $J_{1c}=J_{2c}$, if we take the chemical potentials for both junctions are identically the same,
  we are then led to the same value of the phase difference, {\it i.e.}, $\gamma_1=\gamma_2$.  In this case the corresponding SQUID is a trivial one, which is just double of a single Josephson junction on each side; the magnetic flux, $\Phi=\gamma_2-\gamma_1=0$,\footnote{ We have set the fluxoid number $n=0$, this is because our numerics can only be performed in the vicinity of $\Phi=0$. Please see the discussions below.} vanishes, and there is no interference between the two junctions.

\begin{figure}[h]
   \includegraphics[trim=0cm 0cm 0cm 0cm, clip=true,scale=0.55]{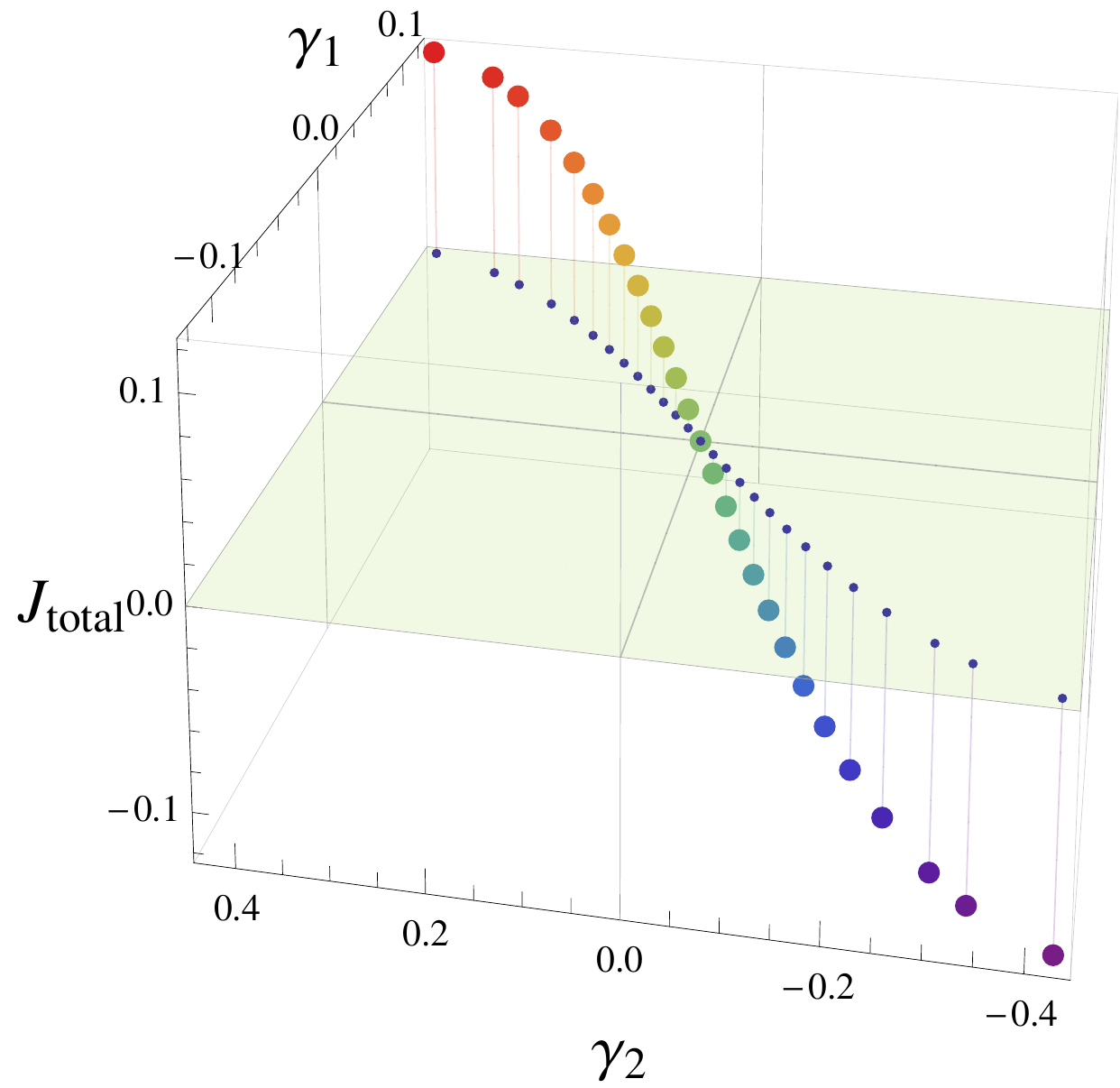}
    \includegraphics[trim=0cm 0cm 0cm 0cm, clip=true,scale=0.55]{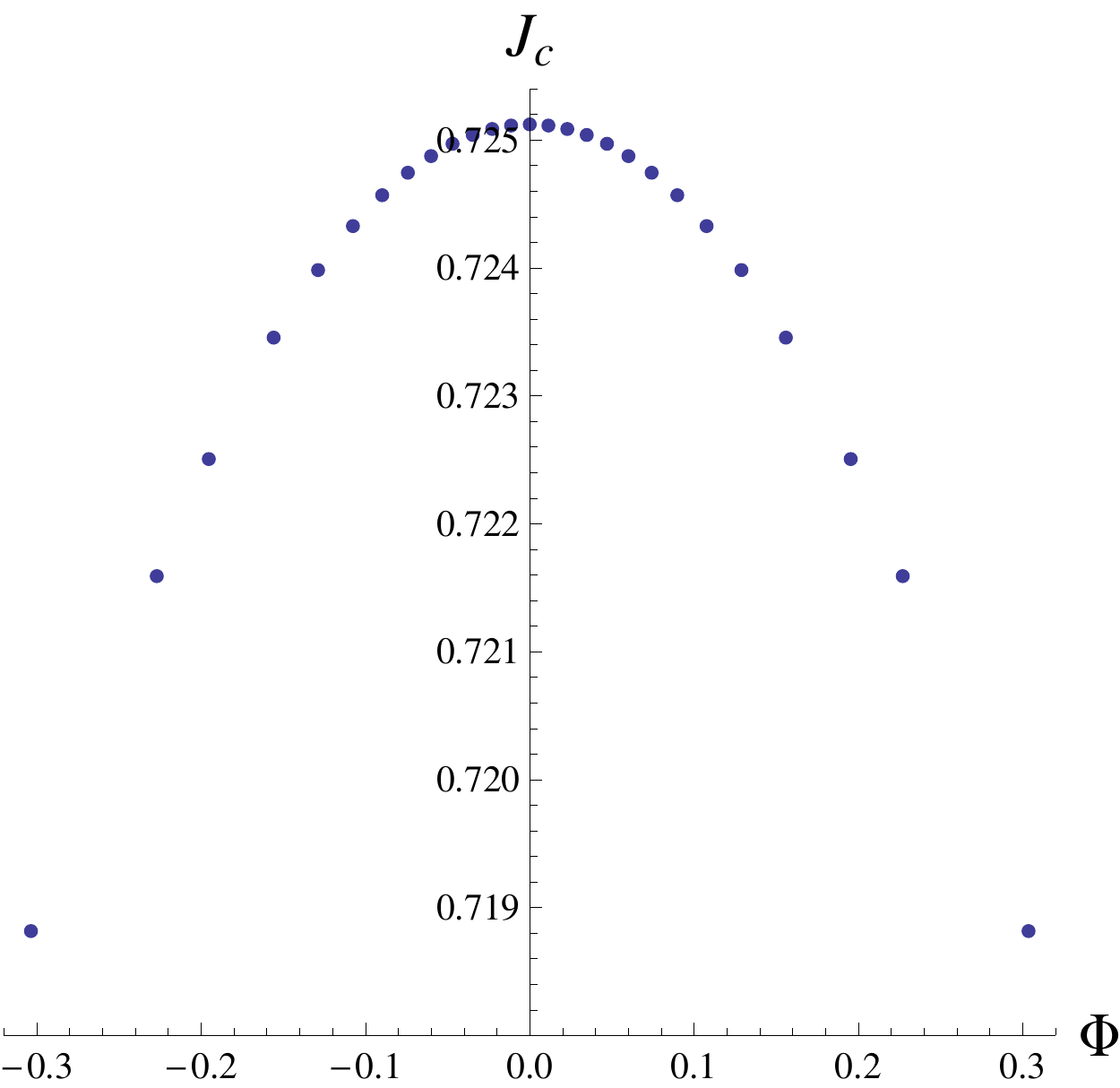}
  \caption{\label{Jtotal} (Left.) The total current $J_{total}$ versus phase difference $\gamma_1$ and $\gamma_2$. The larger points represent $J_{total}$ while the smaller points are the projections of the larger points on $J_{total}=0$ plane; (Right.) The maximal current $J_c$ versus the magnetic flux $\Phi$ from the relation \eqref{jsin}.}
\end{figure}

To overcome this trivial situation, we set the chemical potentials different on two sides of the ring as shown in Fig.\ref{mu}, and then we can get a non-trivial interference between the two junctions, because in this case the phase differences are different for two junctions.  Below we will work in this spirit and manage to get the general relations between the maximal current and the magnetic flux $\Phi$ in the general setup.
The parameters we choose are like those in Fig.\ref{mu}. The input values of the current $J_{i}$ run from $-0.06\to0.06$,  and then the total current $J_{total}$ are from $-0.12\to0.12$. By performing the numerical calculations, we will get a list of the phase differences $\gamma_1$ and $\gamma_2$ for two junctions. Because we scan the values of $J_{i}$,  we can obtain a one parameter curve in the 3D space spanned by $(\gamma_1, \gamma_2, J_{totoal})$, which is plotted in the left panel of Fig.\ref{Jtotal}.
In order to obtain the interference relation between $J_{total}$ and the phase differences $\gamma_i$ ($i=1,2$), we fit the sine relations $J_i=J_{ic}\sin(\gamma_i)$ for two junctions $i$  separately, and then $J_{total}=J_1+J_2$. The fitted result is
\be J_{total}=0.541\sin(\gamma_1)+0.184\sin(\gamma_2). \ee
 Here $J_{1c}=0.541$ and $J_{2c}=0.184$, clearly they are not equal. When $J_{1c}\neq J_{2c}$,  the general form of Eq.\eqref{j1j2} is
\be\label{jtotal2}
J_{total}=J_{1c}\sin(\gamma_1)+J_{2c}\sin(\gamma_2)=J_c\sin(\gamma_c),
\ee
where
\be
J_c&=&\sqrt{J_{1c}^2+J_{2c}^2+2J_{1c}J_{2c}\cos(\gamma_2-\gamma_1)},\label{jsin}\\
\gamma_c&=&\gamma_1+\arctan\left(\frac{J_{2c}\sin(\gamma_2-\gamma_1)}{J_{1c}+J_{2c}\cos(\gamma_2-\gamma_1)}\right)
\nno\\&&+\bigg\{\begin{array}{c}
0,\quad \text{if}\quad J_{1c}+J_{2c}\cos(\gamma_2-\gamma_1)\geq0 \\
\pi,\quad \text{if}\quad J_{1c}+J_{2c}\cos(\gamma_2-\gamma_1)<0
\end{array}
.
\ee
Note that in the parameter range we choose, $J_{1c}+J_{c2}\cos(\gamma_2-\gamma_1)\geq0$ is always satisfied. We therefore have  $\gamma_c=\gamma_1+\arctan\left[{J_{2c}\sin(\gamma_2-\gamma_1)}/({J_{1c}+J_{2c}\cos(\gamma_2-\gamma_1)})\right]$.
%{\blue although it is not important to the problem we are considering}.
Notice again that we have $\gamma_2-\gamma_1=\Phi$ in our model. By virtue of Eq.\eqref{jsin}, we  plot the relation between the maximal current $J_c$ and the magnetic flux $\Phi$ in the right panel of Fig.\ref{Jtotal}.  Here we cannot produce a complete periodic behavior of $J_c$ with respect to the magnetic flux $\Phi$, but only a part of a period. The reason is that our numerical calculations are done
 in the vicinity of $\nu(\chi)=0$, that is, for small values of $\nu(\chi) $. This is caused by the numerical methods we used. For higher values of $J$ or $\nu(\chi)$, the numerical stability and the numerical precision are out of control. A similar situation also appears in the study of holographic Josephson junctions~\cite{Horowitz:2011dz,Wang:2011rva,Wang:2011ri,Wang:2012yj}

 \section{Conclusions}
 \label{sect:con}
 We constructed a holographic model of SQUID in the Einstein-Maxwell-complex scalar theory with a negative cosmological constant by compactifying one spatial direction of the Schwarzschild-AdS black brane. A general relation between the maximal current and the magnetic flux through the SQUID ring was deduced via numerical methods. We worked with a chemical potential so that there are only different depths of the chemical potential for the two junctions. But other differences in the chemical potential for the two junctions will lead to a similar result. Note that the probe limit was adopted in this paper, it is therefore of interest to  study the effect of back reaction of  the matter fields. In addition, considering the two junctions we studied here are the SNS form, it would be interesting to discuss the case of the SQUID ring composed by two superconductor-insulator-superconductor (SIS) junctions.

\section*{Acknowledgments}
 We would like to thank Shingo Takeuchi and Li-Fang Li for their work in the early stage of this paper; H.Q.Z. would like to thank Antonio M. Garc\'ia-Garc\'ia for his helpful comments.
R.G.C. was supported in part by the National Natural Science Foundation of China
(No.10821504, No.11035008); Y.Q.W. was supported by the National Natural Science Foundation of China
(No.11005054);
 H.Q.Z. was supported by a Marie Curie International Reintegration Grant PIRG07-GA-2010-268172.

%\section*{References}
%\begin{multicols}{1}

%\end{multicols}

 % \end{CJK*}
\end{document}